\pdfoutput=1
\RequirePackage{ifpdf}
\ifpdf % We are running pdfTeX in pdf mode
\documentclass[pdftex]{sigma}
\else
\documentclass{sigma}
\fi

\begin{document}

\allowdisplaybreaks

\renewcommand{\PaperNumber}{113}

\FirstPageHeading

\ShortArticleName{Breaking Pseudo-Rotational Symmetry}

\ArticleName{Breaking Pseudo-Rotational Symmetry through $\boldsymbol{{\bf H}^2_+}$ Metric Deformation in the Eckart Potential Problem}

\Author{Nehemias LEIJA-MARTINEZ~$^\dag$, David Edwin ALVAREZ-CASTILLO~$^\ddag$\\ and Mariana KIRCHBACH~$^\dag$}

\AuthorNameForHeading{N.~Leija-Martinez, D.E.~Alvarez-Castillo and M.~Kirchbach}

\Address{$^\dag$~Institute of Physics, Autonomous University of San Luis Potosi,\\
\hphantom{$^\dag$}~Av. Manuel Nava 6, San Luis Potosi, S.L.P. 78290, Mexico}
\EmailD{\href{mailto:nemy@ifisica.uaslp.mx}{nemy@ifisica.uaslp.mx}, \href{mailto:mariana@ifisica.uaslp.mx}{mariana@ifisica.uaslp.mx}}

\Address{$^\ddag$~H.~Niewodnicza\'nski Institute of Nuclear Physics, Radzikowskiego 152, 31-342 Krak\'ow, Poland}
\EmailD{\href{mailto:david.alvarez@ifj.edu.pl}{david.alvarez@ifj.edu.pl}}

\ArticleDates{Received October 12, 2011, in f\/inal form December 08, 2011; Published online December 11, 2011}

\Abstract{The peculiarity of the Eckart potential problem
on {\bf H}$^2_+$ (the upper sheet of the two-sheeted two-dimensional
hyperboloid),
to preserve the $(2l+1)$-fold degeneracy of the states
typical for the geodesic motion there, is usually explained in casting
the respective Hamiltonian in terms of the Casimir invariant of
an so(2,1) algebra, referred to as potential algebra. In general,
there are many possible
similarity transformations of the symmetry algebras of the free motions on
curved surfaces towards potential algebras, which are not all
necessarily unitary.
In the literature, a transformation of the symmetry algebra of the geodesic
motion on {\bf H}$_+^2$ towards the potential algebra of Eckart's Hamiltonian
has been constructed for the prime purpose to prove that
the Eckart interaction belongs to the class of Natanzon potentials.
We here take a dif\/ferent path and search for a transformation which
connects the $(2l+1)$ dimensional representation space of the
pseudo-rotational so(2,1) algebra, spanned by the rank-$l$
pseudo-spherical harmonics, to the representation space of equal dimension
of the potential algebra and f\/ind a transformation of the scaling type.
Our case is that in so doing one is producing a deformed isometry copy
to ${\mathbf H}^2_+$ such that the free motion on the copy
is equivalent to a motion on ${\mathbf H}^2_+ $, perturbed by a
$\coth$ interaction.
In this way, we link the so(2,1) potential algebra concept of the
Eckart Hamiltonian to a~subtle type of pseudo-rotational symmetry breaking
through ${\mathbf H}^2_+$ metric deformation.
From a~technical point of view, the results reported here are obtained by
virtue of certain nonlinear f\/inite expansions of Jacobi polynomials into
pseudo-spherical harmonics. In due places, the pseudo-rotational
case is paralleled by its so(3) compact analogue, the cotangent
perturbed motion on S$^2$.
We expect awareness of dif\/ferent so(2,1)/so(3) isometry copies
to benef\/it simulation studies on curved manifolds of many-body systems.}


\Classification{47E05; 81R40}

\section{Introduction}

Group theoretical approaches to bound state problems in physics have
played a pivotal r\'ole in our understanding of
spectra classif\/ications. It is a well known fact, that several of
the exactly solvable quantum mechanical potentials give rise to spectra which
fall into the irreducible representations of certain Lie groups.
As a representative case, we wish to mention the widely studied class of
Natanzon potentials \cite{Natanzon} known to produce spectra that
populate multiplets of the pseudo-rotational algebras so(2,2)/so(2,1).
This phenomenon has been well understood in casting the Hamiltonians
of the potentials under consideration as
Casimir invariants of so(2,2)/so(2,1) algebras in representations
not necessarily unitarily equivalent to the pseudo-rotational and
referred to as
``potential algebras''~\cite{Alhassid,Quesne}.
One popular representative of the
Natanzon class potentials is the Eckart potential, suggested
by Manning and Rosen~\cite{Manning_Rosen}
for the description of the vibrational modes of
diatomic molecules. Its so(2,2)/so(2,1) potential algebras have been
constructed, among others, in~\cite{WuAlhassid,Levai,Salamo,Salamo2,AsimG,AsimG_2}.
One can start with considering the
Eckart potential on {\bf H}$^2$, the two-sheeted
2D hyperbolic surface, def\/ined as
\begin{gather*}
{\mathbf H}^2_\pm : \ z^2 -x^2-y^2=R^2,
%\label{H2}
\end{gather*}
where $R$ is a constant.
This geometry has been studied in great detail~\cite{Willard}, and
f\/inds applications, among others,
as a coherent state manifold \cite{book,Bogdanova}.
One usually chooses the upper sheet~${\mathbf H}^2_+$,
 which corresponds to the following parametrization in global coordinates
\begin{gather*}
z=R\cosh \eta, \qquad x=R\sinh \eta \cos \varphi,\nonumber\\
y = R\sinh \eta \sin\varphi, \qquad z>\sqrt{x^2+y^2}, \qquad
\eta \in [0,\infty).
%\label{parametrization}
\end{gather*}
The free geodesic motion on ${\mathbf H}^2_+$ is now determined by
the eigenvalue problem of the squared pseudo-angular momentum operator~${\mathcal C}$
\begin{gather}
{\mathcal H}{\mathcal Y}_l^m(\eta,\varphi) =
-\frac{\hbar^2}{2M}l(l+1)
{\mathcal Y}_l^m(\eta, \varphi),\qquad {\mathcal H}=-
\frac{\hbar^2}{2M}{\mathcal C},\nonumber\\
{\mathcal C} =
\frac{1}{\sinh \eta } \frac{\partial }{\partial \eta }
\sinh \eta \frac{\partial }{\partial \eta } + \frac{1}{\sinh^2 \eta }
\frac{\partial ^2}{\partial \varphi^2},
\label{Wu1}
\end{gather}
so that the symmetry of the Hamiltonian, ${\mathcal H}$, is so(2,1).
Here, ${\mathcal Y}_l^m(\eta ,\varphi)$ are the standard
pseudo-spherical harmonics \cite{Kim_Noz}
\begin{gather}
{\mathcal Y}_l^m(\eta ,\varphi) = P_l^m(\cosh \eta )e^{im\varphi},\qquad
\eta \in (-\infty, +\infty),
\label{psdsfr}
\end{gather}
with $P_l^m(\cosh\eta)$ being the associated Legendre
functions of a hyperbolic cosines argument.
In~\cite{AsimG} it has been shown that
the following particular similarity transformation of~${\mathcal C}$
 \begin{gather}
{\mathbf F}(\eta ){\mathcal C}{\mathbf F}^{-1}(\eta ) =
\widetilde{\mathcal C}, \qquad
 {\mathbf F}(\eta)=\sqrt{\frac{\sinh 2g(r)}{g^\prime (r)}},
\qquad \tanh^2g(r)=z=e^{-r},
\label{W2}
\end{gather}
transforms equation~(\ref{Wu1}) into a Natanzon equation in the
 $z$ variable, and to the central Eckart potential problem in ordinary 3D
f\/lat position space in the $r$ variable.
In this fashion it became possible to cast the Schr\"odinger Hamilton operator
with the Eckart potential, treated as a central interaction, in the form of
a particular Casimir invariant of the~so(2,1) algebra.

Compared to \cite{WuAlhassid}, and \cite{AsimG},
we here exclusively focus on the
relationship between the free, and the $\coth\eta $ perturbed motions on
{\bf H}$_+^2$ and f\/ind a transformation that
connects the $({\mathcal C}+2b\coth\eta )$- and
${\mathcal C}$-eigenvalue problems and their respective invariant spaces,
a transformation which has not been reported in the literature so far.
The ${\mathcal C}$ invariant spaces are $(2l+1)$-dimensional, non-unitary,
and spanned by the pseudo-spherical harmonics in~(\ref{psdsfr}).
Towards our goal, we develop a~technique for solving
the $\left( {\mathcal C}+2b\coth\eta \right)$-eigenvalue problem
which slightly dif\/fers from the one of standard use.
Ordinarily, the respective 1D Schr\"odinger equation is resolved by
reducing it to the hyper-geometric dif\/ferential equation satisf\/ied by
the Jacobi polynomials. We here instead construct the solutions directly
in the basis of the free geodesic motion, i.e.\ in the
${\mathcal Y}_l^m(\eta,\varphi)$ basis.
In so doing, we f\/ind superpositions of exponentially damped
pseudo-spherical harmonics
which describe the perturbance by a hyperbolic cotangent function
of the free geodesic motion on {\bf H}$_+^2$. The
matrix similarity transformation
between the ${\mathcal C}$ and $({\mathcal C}+2b\coth\eta) $
eigenvalue problems is then read of\/f from
those decompositions which allow
to interpret the Hamiltonian of
the perturbed motion as a Casimir invariant of an
so(2,1) algebra in a representation unitarily nonequivalent to the
pseudo-rotational.
The consequence is a breakdown of the pseudo-rotational
invariance of the free geodesic motion only at
the level of the representation functions
which becomes apparent through a deformation of the
metric of the hyperbolic space, and
without a breakdown of the $(2l+1)$-fold degeneracy patterns
in the spectrum.
Moreover, due to cancellation ef\/fects occurring by
virtue of specif\/ic recurrence relations among
associated Le\-gendre functions, the similarity
transformation presents itself
simple and, as explained in the concluding summary section,
may specif\/ically benef\/it applications in problems with an
appro\-xi\-mate so(2,1) symmetry.
The problem under investigation is well known to transform into the
Rosen--Morse potential problem on S$^2$ by a complexif\/ication of
the hyperbolic angle, followed by a complexif\/ication of the
strength of the $\coth\eta $ potential. On this basis, the statement on the
symmetry and degeneracy properties of the Eckart potential on~{\bf H}$_+^2$
can easily be transferred to the trigonometric
Rosen--Morse potential problem on~S$^2$.

The paper is structured as follows.
In the next section we present the eigenvalue problem of the
$\coth$-perturbed particle motion on~{\bf H}$_+^2$.
There, we furthermore work out the respective solutions as expansions
in the basis of the standard pseudo-spherical harmonics,
f\/ind the non-unitary transformation
relating the free and the perturbed geodesic motions, present the class of recurrence relations among
associated Legendre functions that triggers the above similarity
transformation. In due places, the conclusions regarding the
Rosen--Morse potential problem on~S$^2$ have been drawn in parallel
to those regarding the Eckart potential.
The paper ends with concise summary and conclusions.

\section[Scalar particle motions on H$^2_+$ and S$^2$
and representations for the so(2,1) and so(3) algebras
 unitarily nonequivalent to the respective pseudo-rotational
 and rotational ones]{Scalar particle motions on $\boldsymbol{{\bf H}^2_+}$ and $\boldsymbol{{\rm S}^2}$
and representations\\ for the so(2,1) and so(3) algebras
 unitarily nonequivalent\\ to the respective pseudo-rotational
 and rotational ones}\label{section2}

\subsection{From free to perturbed motions}

\begin{figure}[t]
\centering
{\includegraphics[width=7.75cm]{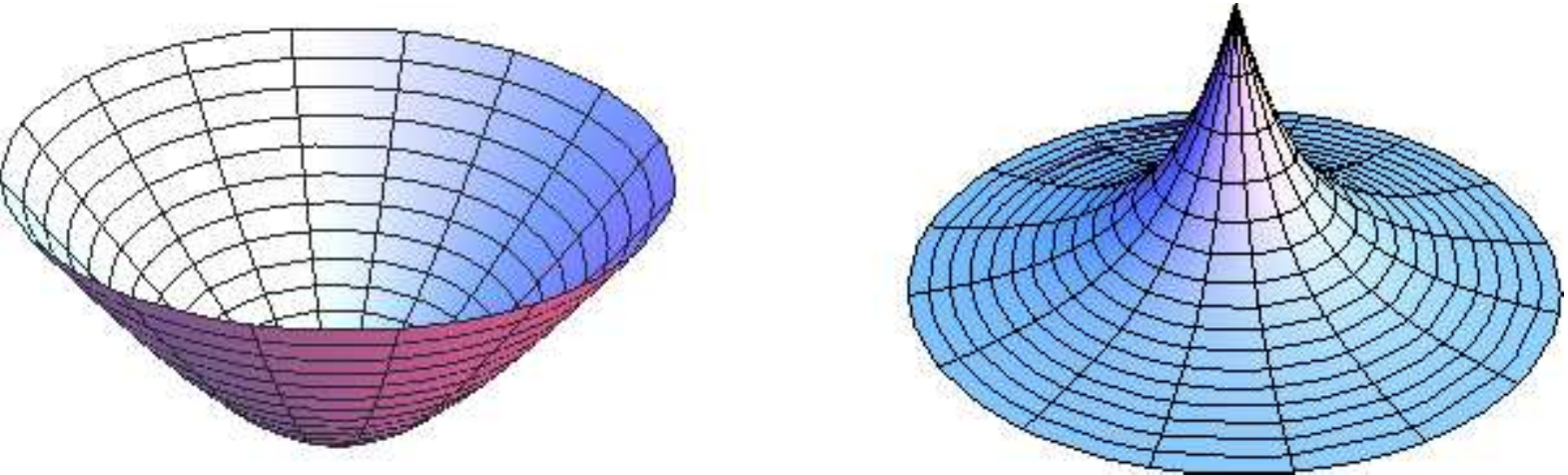}}
\caption{Breaking of the pseudo-rotational symmetry at the
level of the metric.
A regular {\bf H}$_+^2$ hyperboloid associated with the so(1,2) scalar,
$(z^2-x^2-y^2)$ (left), and its
non-unitary deformation,
$\exp (-\alpha_{(l=0)}\eta /2)(z^2-x^2-y^2)$
(right) with $\alpha_{(l=0)}=4b$, and $\eta =\coth^{-1}\frac{z}{\sqrt{x^2+y^2}}$.
Schematic presentation for $b=1$.\label{tld2}}
\end{figure}

We begin with taking a closer look on equation~(\ref{Wu1}) for the
particular case of $l=0$
\begin{gather}
-\frac{\hbar^2}{2M}{\mathcal C}{\mathcal Y}_0^0(\eta,\varphi ) = 0,
\label{Wu33}
\end{gather}
and subject it to a similarity transformation to obtain
\begin{gather}
-\frac{\hbar^2}{2M}\left[{\mathbf F}^{-1}(\eta){\mathcal C}
{\mathbf F}(\eta)\right] \left[
 {\mathbf F}^{-1}(\eta) {\mathcal Y}_0^0(\eta,\varphi )\right] =
-\frac{\hbar^2}{2M}{\widetilde {\mathcal C}}
\left[ {\mathbf F}^{-1}(\eta) {\mathcal Y}_0^0(\eta,\varphi )\right]=0,
\label{Wu44}
\end{gather}
with
$\widetilde{\mathcal C}={\mathbf F}^{-1}(\eta ){\mathcal C}{\mathbf F}(\eta )$.
The general expression for any transformation
in (\ref{Wu44}) is easily worked out (cf.~\cite{AsimG}) and reads
\begin{gather*}
\widetilde{\mathcal C}={\mathbf F}^{-1}(\eta ){\mathcal C}{\mathbf F}(\eta )
 =
{\mathcal C} +\frac{1}{{\mathbf F}(\eta )}\frac{{\mathrm d}^2
{\mathbf F}(\eta )}{{\mathrm d}^2\eta}
+2\frac{1}{{\mathbf F}(\eta )}
\frac{{\mathrm d}{\mathbf F}(\eta )}{{\mathrm d}\eta }
\left( \frac{{\mathrm d}}{{\mathrm d}\eta }+\frac{1}{2}\coth \eta \right).
%\label{Wu55}
\end{gather*}
We here chose ${\mathbf F}(\eta)$ dif\/ferently than in (\ref{W2})
and consider the following scaling similarity transformation
\begin{gather}
{\mathbf F}(\eta )=e^{\frac{\alpha_{(l=0)} \eta }{2}}, \qquad
\alpha_{(l=0)}=4b.
\label{Wu4}
\end{gather}
Substitution of (\ref{Wu4}) into (\ref{Wu44}) amounts to
\begin{gather}
-\frac{\hbar^2}{2M}{\widetilde C}{\widetilde{\mathcal Y}}_0^0(\eta,\varphi) =
-\frac{\hbar^2}{2M}\left[
e^{-\frac{\alpha_{(l=0)}\eta}{2}}{\mathcal C}
e^{\frac{\alpha_{(l=0)}\eta}{2}}\right]\,
\left[ e^{-\frac{\alpha_{(l=0)}\eta}{2}}{\mathcal Y}_0^0(\eta,\varphi)\right]\nonumber\\
\phantom{-\frac{\hbar^2}{2M}{\widetilde C}{\widetilde{\mathcal Y}}_0^0(\eta,\varphi)}{}  =
-\frac{\hbar^2}{2M}\left( {\mathcal C} -\frac{\alpha_{(l=0)}^2}{4} +2b\coth\eta
\right){\widetilde {\mathcal Y}_0^0}(\eta, \varphi)=0,\nonumber\\
{\widetilde {\mathcal Y}_0^0}(\eta,\varphi)  =
e^{-\frac{\alpha_{(l=0)}\eta}{2}}{\mathcal Y}_0^0(\eta, \varphi),
\label{Wuuuu}
\end{gather}
which turns equivalent to the standard form of the
$\coth \eta $ perturbed motion on {\bf H}$_+^2$, for the ground state
\begin{gather}
-\frac{\hbar^2}{2M}\left( {\mathcal C}+2b \coth\eta \right)
{\widetilde {\mathcal Y}_0^0}(\eta,\varphi)
=-\frac{\hbar^2}{2M}4b^2 {\widetilde {\mathcal Y}_0^0}(\eta,\varphi),
\label{Wu5}
\end{gather}
where the special notation, ${\widetilde {\mathcal Y}_0^0}(\eta, \varphi)$,
has been introduced in (\ref{Wuuuu}) for
the exponentially damped pseudo-spherical harmonic under consideration.
We now notice that ${\mathcal C}$ in~(\ref{Wu33}) is the Casimir invariant of
the so(1,2) isometry
algebra of the ${\mathbf H}^2$ hyperboloid,
described by the $|{\mathcal Y}_0^0(\eta,\varphi)|$
unit surface, displayed in the l.h.s.\ of
Fig.~\ref{tld2}, and that, correspondingly,
${\widetilde {\mathcal C}}=e^{-\frac{\alpha_{(l=0)}\eta}{2}}
{\mathcal C}e^{\frac{\alpha_{(l=0)}\eta}{2}}$ in~(\ref{Wuuuu})
is same on the deformed
surface $e^{-\alpha_{(l=0)}\eta /2}|{\mathcal Y}_0^0(\eta,\varphi)|$,
shown on the r.h.s.\ in Fig.~\ref{tld2}.

From this perspective, equation (\ref{Wuuuu})
shows that the free motion on the deformed metric in Fig.~\ref{tld2}
is equivalent to
the $\coth \eta $ perturbed motion on ${\mathbf H}^2_+$.
The rest of the article is devoted to generalize equation~(\ref{Wuuuu}) to higher
$l$ values.
From now onwards we shall switch to dimensionless units, $\hbar=1$, $2M=1$,
for the sake of simplicity.

\begin{figure}[t]\centering
\includegraphics[height=35mm]{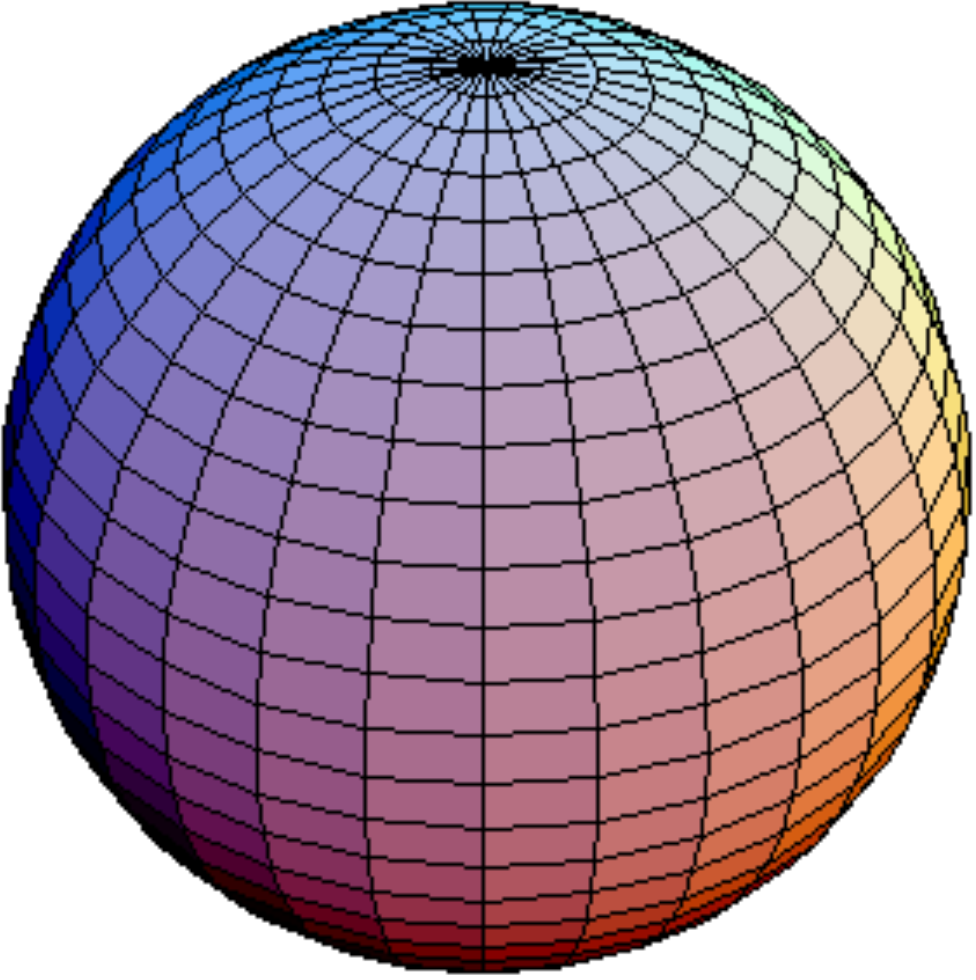} \qquad \qquad \qquad \includegraphics[height=35mm]{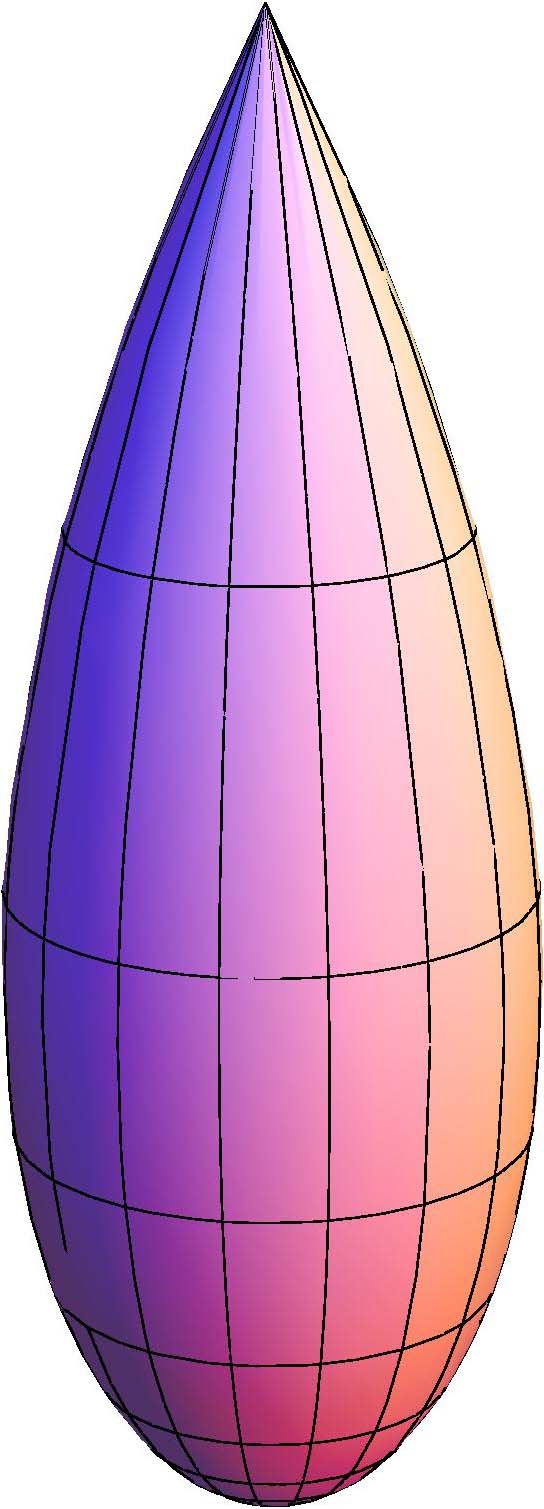}
\caption{The S$^2$ spherical metric, $|Y_{00}(\theta, \varphi)|$
(left) in comparison to the exponentially deformed one,
$|\widetilde {Y}_{00}(\theta,\varphi )|$ (right) for b=1.
The cotangent perturbed rigid rotor on S$^2$ is equivalent
(up to a shift by constant) to free motion on the deformed metric, as visible
from equation~(\ref{Lu1}).}
\label{baseball}
\end{figure}

In parallel, we notice that as long as so(1,2), and so(3) are
related by a Wigner rotation,
${\mathcal C}$~in equation~(\ref{Wu1})
is related to the well known
expression of the
standard squared orbital angular momentum ${\mathbf L}^2$
\begin{gather*}
{\mathbf L}^2=
-\frac{1}{\sin\theta }\frac{\partial }{\partial \theta }\sin \theta
\frac{\partial }{\partial \theta } -\frac{1}{\sin^2\theta }
\frac{\partial^2}{\partial\varphi^2},
%\label{oam}
\end{gather*}
by a complexif\/ication of the polar angle
\begin{gather*}
{\mathbf L}^2\stackrel{\theta \to i\eta }{\longrightarrow}{\mathcal C}.
%\label{cmplxfctn}
\end{gather*}
It is the type of complexif\/ication that
takes the hyperboloid {\bf H}$^2_+$ to
the sphere~S$^2$.
Correspon\-dingly, the scaling transformation in equation~(\ref{Wu4})
will take the free geodesic motion on S$^2$ to the one perturbed by the
trigonometric Rosen--Morse potential there
\begin{gather}
{\widetilde {\mathbf L}^2}\widetilde{Y}_0^0(\theta,\varphi) =
\left( {\mathbf L}^2-2b \cot\theta -\frac{\alpha_{(l=0)}^2}{4} \right)
\widetilde{Y}_0^0(\theta,\varphi)=0, \nonumber\\
\widetilde{Y}_0^0(\theta,\varphi) =
e^{-\frac{\alpha_{(l=0)}\theta}{2}}Y_0^0(\theta,\varphi).
\label{Lu1}
\end{gather}
The counterpart of Fig.~\ref{tld2} is displayed on Fig.~\ref{baseball}.

\subsection{The perturbed geodesic motions}

We now consider the general case of a perturbance of the free geodesic motion
in (\ref{Wu1}) by a~hyperbolic cotangent potential
\begin{gather}
\left[ {\mathcal C}+2b\coth \eta\right]{\mathcal X}(\eta, \varphi) \nonumber\\
\qquad{} =
\left[ \frac{1}{\sinh \eta } \frac{\partial }{\partial \eta }
\sinh \eta \frac{\partial }{\partial \eta } +\frac{1}{\sinh^2 \eta }
\frac{\partial^2}{\partial\varphi^2}
 + 2b\coth \eta\right]{\mathcal X}(\eta ,\varphi)
 =
-\epsilon {\mathcal X}(\eta, \varphi),
\label{Gl3}\\
{\mathcal X}(\eta ,\varphi)=e^{-\frac{\alpha\eta }{2}}F(\eta )e^{i{\widetilde m}\varphi },
\label{Gl3_Anh}
\end{gather}
where $\epsilon$ stands for the energy in dimensionless units.
The $\eta $ and $\varphi$ dependencies of the eigenfunctions
separate, and we admitted for the possibility that ${\widetilde m}$
 may not take same value as
the magnetic quantum number in the free case.
In what follows we attach indices to ${\mathcal X}(\eta, \varphi)$
as ${\mathcal X}(\eta, \varphi)\longrightarrow
{\mathcal X}_l^{\widetilde m}(\eta, \varphi)$,
and search for
$({\mathcal C}+2b\coth\eta) $ eigenfunctions
in the form of superpositions of damped
pseudo-spherical harmonics,
${\widetilde {\mathcal Y}}_l^m(\eta ,\varphi)$, with
$m\in \lbrack{\widetilde m}, l\rbrack $, according to
\begin{gather}
 {\mathcal X}_l^{\widetilde m}(\eta, \varphi) =
\left[ \sum_{m={\widetilde m}}^{m=l}a_{({\widetilde m}+1)(m+1)}^l e^{-im\varphi}
{\widetilde {\mathcal Y}}_l^m(\eta, \varphi)\right]
e^{i{\widetilde m}\varphi },\nonumber\\
{\widetilde {\mathcal Y}}_l^m(\eta ,\varphi)  =
e^{-\frac{\alpha\eta }{2}}{\mathcal Y}_l^m(\eta ,\varphi).
\label{decomposition}
\end{gather}
 Furthermore,
$a_{({\widetilde m}+1)(m+1)}^le^{i({\widetilde m}-m)\varphi} $
(with constant $a_{({\widetilde m}+1)(m+1)}^l$) are elements of an
$(l+1)\times (l+1)$ dimensional matrix.
We begin with the observation that upon substituting (\ref{decomposition})
in (\ref{Gl3}),
the particle motion on {\bf H}$_+^2$, hindered by a $\coth \eta$
interaction, equivalently rewrites to
\begin{gather}
e^{-\frac{\alpha\eta }{2}}
\left[ \frac{1}{\sinh \eta}\frac{\partial }{\partial \eta }
\sinh \eta \frac{\partial }{\partial \eta } -\frac{{\widetilde m}^2}
{\sinh^2 \eta }
+ \frac{\alpha ^2}{4}
+\alpha D
\right]
\left[\sum _{|m|=|{\widetilde m}|}^l a^l_{(|{\widetilde m }|+1)(|m|+1)}
P_l^{|m|}(\cosh \eta)\right]
e^{i{\widetilde m}\varphi}\nonumber\\
 = -\epsilon e^{-\frac{\alpha\eta }{2}}\left[\sum _{|m|=|{\widetilde m}|}^l a^l_{(|{\widetilde m }|+1)(|m|+1)}
P_l^{|m|}(\cosh \eta)\right]
e^{i{\widetilde m}\varphi}, \quad
D=\left( \frac{2b}{\alpha }-\frac{1 }{2}\right)\coth \eta
- \frac{\partial }{\partial \eta},
\label{insight}
\end{gather}
where we dragged the exponential function through ${\mathcal C}$
from the very right to the very left, and inserted equation~(\ref{decomposition})
for ${\mathcal X}_l^{\widetilde m}(\eta, \varphi)$.
Next we consider the particular case when the sum in
equation~(\ref{decomposition})
contains one term only, i.e.\ when ${\widetilde m}=l$:
\begin{gather*}
  {\mathcal X}_l^l(\eta, \varphi)
=a^l_{(l+1),(l+1)}e^{-\frac{\alpha\eta}{2}}P_l^l(\cosh\eta)e^{il\varphi }.
%\label{max_m}
\end{gather*}
In this case, the part of~(\ref{insight}) behind the exponential
would reduce to the dif\/ferential equation
for the associated Legendre
functions, shifted by the constant, $\alpha^2/4$, provided
\begin{gather}
DP_l^l(\cosh\eta)=0, \qquad P_l^l(\cosh\eta )=\sinh^l\eta,
\label{alfa_fix}
\end{gather}
were to hold valid.
The latter equation imposes the following condition
 on $\alpha$
\begin{gather}
\alpha\longrightarrow \alpha_l, \qquad
\alpha_l=\frac{2b}{l+\frac{1}{2}},
\label{value_alfa}
\end{gather}
where we re-labeled $\alpha$ to $\alpha\to \alpha_l$.
 In ef\/fect, the $({\mathcal C}+2b\coth \eta)$ eigenvalue problem
simplif\/ies to the ${\mathcal C}$
eigenvalue problem subjected to a scaling similarity
transformation according to
\begin{gather}
\left[ {\mathcal C}+2b\coth \eta\right]
{\mathcal X}_l^l(\eta, \varphi)=
\left[
e^{-\frac{\alpha_l\eta }{2}}\left( {\mathcal C}+\frac{\alpha_l^2}{4}\right)
e^{\frac{\alpha_l\eta }{2}}\right]{\mathcal X}_l^l(\eta, \varphi)
=
\left( l(l+1) +\frac{\alpha_l^2}{4}\right){\mathcal X}_l^l(\eta, \varphi)
,\nonumber\\
 {\mathcal X}_l^l(\eta, \varphi)
=
a^l_{(l+1),(l+1)}e^{-\frac{\alpha_l\eta }{2}}{\mathcal Y}_l^l(\eta, \varphi).
\label{scaling_1}
\end{gather}
We set $a^l_{(l+1),(l+1)}=1$ for simplicity.
As a next step, we consider the case of a two-term decomposition
in (\ref{decomposition}), when
\begin{gather*}
{\mathcal X}_{l}^{l-1}(\eta, \varphi )=
e^{-\frac{\alpha_l\eta }{2}}
\left( a^l_{ll}P_{l}^{l-1}(\cosh\eta) +
a^l_{(l)(l+1)}P_l^l(\cosh \eta )\right)e^{i(l-1)\varphi}.
%\label{exmpl_1}
\end{gather*}
Substituting in (\ref{Gl3}), and dragging the exponential
factor again from the very right to the very left, amounts to
\begin{gather}
\left[{\mathcal C}+2b\coth\eta\right]{\mathcal X}_l^{l-1}(\eta ,\varphi) =
e^{-\frac{\alpha_l\eta }{2}}
\left[ \frac{1}{\sinh \eta}\frac{\partial }{\partial \eta }
\sinh \eta \frac{\partial }{\partial \eta } -\frac{(l-1)^2}
{\sinh^2 \eta }
+ \frac{\alpha ^2_l}{4}
+\alpha_l D_l
\right]\nonumber\\
\qquad\quad{}
\times\Big( a^l_{ll}P_l^{l-1}(\cosh \eta)
+ a^l_{(l)(l+1)}P_l^{l}(\cosh \eta)
 \Big)
e^{i(l-1)\varphi}\nonumber\\
\qquad{}
=\left( l(l+1)+\frac{\alpha_l^2}{4}\right)\!e^{-\frac{\alpha\eta }{2}}
\Big(a^l_{ll}P_l^{l-1}(\cosh \eta)
+ a^l_{(l)(l+1)}P_l^{l}(\cosh \eta)
\Big)e^{i(l-1)\varphi}\!,
\nonumber\\
D_l = \left( l\coth \eta -
\frac{\partial }{\partial \eta}\right),
\label{insight_2}
\end{gather}
where the expression for $D_l$, corresponding to $D$ from (\ref{insight})
with an attached label $\ell$, i.e.\ $D\to D_l$,
has been obtained in making use use of (\ref{value_alfa}).
The unknown constant, $a^l_{(l)(l+1)}$, is now determined from the condition
\begin{gather*}
-\frac{(l-1)^2}{\sinh^2\eta}a_{(l)(l+1)}^l P_l^l(\cosh\eta) +
\alpha_lD_la_{ll}^lP_l^{l-1}(\cosh \eta )=-\frac{l^2}{\sinh^2\eta }
a_{(l)(l+1)}^lP_l^l(\cosh\eta),
%\label{cond_2}
\end{gather*}
setting once again $a^l_{ll}=1$.
Taking into account that $P_l^{l-1}(\cosh\eta )=\sinh^{l-1}\eta \cosh\eta$,
one arrives at the constraint
\begin{gather*}
a^l_{(l)(l+1)}=\frac{\alpha_l}{-l^2+(l-1)^2},
\end{gather*}
thus f\/ixing the $a^l_{l(l+1)}$ value.
In consequence, the eigenvalue problem in
equation~(\ref{insight_2}) simplif\/ies to same form as previously
found in equation~(\ref{scaling_1}), namely
\begin{gather*}
\left[{\mathcal C}+2b\coth\eta \right]
{\mathcal X}_l^{l-1}(\eta ,\varphi) =
\left[ e^{-\frac{\alpha_l\eta }{2}}\left( {\mathcal C}+
\frac{\alpha_{l}^2}{4}\right)e^{\frac{\alpha_{l}\eta }{2}}\right]
{\mathcal X}_l^{l-1}(\eta ,\varphi)\nonumber\\
\hphantom{\left[{\mathcal C}+2b\coth\eta \right]{\mathcal X}_l^{l-1}(\eta ,\varphi)}{}
=\left[\widetilde{\mathcal C} +\frac{\alpha^2_{(l)}}{4} \right]
{\mathcal X}_l^{l-1}(\eta,\varphi)  =
\left( l(l+1) +\frac{\alpha _{(l)}^2}{4}\right)
{\mathcal X}_l^{l-1}(\eta, \varphi).
%\label{illustr_2}
\end{gather*}
Proceeding successively in this way, the coef\/f\/icients
$a^l_{({\widetilde m}+1)(m+1)}$ can be found imposing the following condition
\begin{gather}
\left[ -\frac{{\widetilde m}^2}{\sinh^2\eta } +\alpha_{l}D_l \right]
\left[ \sum_{m={\widetilde m}}^{m=l}a_{({\widetilde m}+1)(m+1)}^l e^{-im\varphi}
 {\mathcal Y}_l^m(\eta, \varphi)\right]\nonumber\\
\qquad{}  =
-\left[ \sum_{m={\widetilde m}}^{m=l}a_{({\widetilde m}+1)(m+1)}^l e^{-im\varphi}
\frac{m^2}{\sinh^2\eta } {\mathcal Y}_l^m(\eta, \varphi)\right],
\label{lit}
\end{gather}
guaranteed by virtue of recurrence relations among
associated Legendre functions of the type
\begin{gather}
D_{(l=1)}P_1^0(\cosh\eta ) =\frac{1^2}{\sinh^2\eta } P_1^1(\cosh \eta ),
\label{rcr1}\\
D_{(l=2)}P_2^1(\cosh \eta)=\frac{1}{\sinh^2\eta }P_2^2 (\cosh\eta) ,\\
D_{(l=2)}P_2^0(\cosh\eta )=\frac{2}{3}\frac{1}{\sinh ^2\eta}P_2^1 (\cosh \eta),
 \quad \mbox{etc.}
\label{rcr2}
\end{gather}
In combination with the identity,
$P_l^{|m|}(\cosh\eta )=e^{-im\varphi}{\mathcal Y}_l^m(\eta ,\varphi)$,
equations~(\ref{lit})--(\ref{rcr2}) allow
 to cast the general $({\mathcal C} +2b\coth \eta))
{\mathcal X}_l^{{\widetilde m}}(\eta, \varphi)$
eigenvalue problem for any $l$ and ${\widetilde m}$
into the following equivalent form
\begin{gather}
 \left[{\mathcal C} +2b\coth \eta \right]
{\mathcal X}_l^{{\widetilde m}}(\eta, \varphi)=
\left[ \widetilde { {\mathcal C}}
+ \frac{\alpha_l^2}{4} \right]
{\mathcal X}_l^{{\widetilde m}}(\eta ,\varphi)
 = \left(l(l+1)+\frac{\alpha_l^2}{4}\right)
{\mathcal X}_l ^{{\widetilde m}}(\eta, \varphi),
\label{bibop1}
\end{gather}
with ${\mathcal X}_l^{{\widetilde m}}(\eta, \varphi)$ standing for the
exact solutions of the Eckart potential on {\bf H}$_+^2$ obtained from the
ansatz in~(\ref{decomposition}).
Omitting normalization factors for simplicity,
the eigenfunctions to~(\ref{bibop1}) for some of the lowest~$l$ values,
now cast in matrix form, are calculated as
\begin{gather}
\begin{pmatrix}
 {\mathcal X}_1^0(\eta, \varphi) \\
 {\mathcal X}_1^1(\eta,\varphi)
\end{pmatrix} =
e^{-\frac{2b\eta}{3}}\begin{pmatrix}
1&-\frac{4b}{3}e^{-i\varphi}\\
0&1
\end{pmatrix}
\begin{pmatrix}
{\mathcal Y}_1^0(\eta, \varphi)\\
{\mathcal Y}_1^1(\eta, \varphi)
\end{pmatrix},
\label{bong1}\\
\begin{pmatrix}
{\mathcal X}_2^0(\eta, \varphi)\\
{\mathcal X}_2^1(\eta ,\varphi)\\
{\mathcal X}_2^2(\eta ,\varphi)
\end{pmatrix}
 =
e^{-\frac{2b\eta}{5}}
\begin{pmatrix}
1&-\frac{8b}{15}e^{-i\varphi}&\frac{8b^2}{75}e^{-2i\varphi}\\
0&1& -\frac{4b}{15}e^{-i\varphi}\\
0&0& 1
\end{pmatrix}
\begin{pmatrix}
{\mathcal Y}_2^0(\eta ,\varphi)\\
{\mathcal Y}_2^1(\eta , \varphi)\\
{\mathcal Y}_2^2(\eta ,\varphi)
\end{pmatrix}.
\label{bong}
\end{gather}
Admittedly, the above expressions are easier obtained parting from
the (unnormalized) solutions of equation~(\ref{Gl3}) known from the
literature~\cite{Dutt} to be expressed in terms of Jacobi polynomials,
here denoted by ${\mathcal P}_n^{\gamma, \delta}$, as
\begin{gather}
{\mathcal X}_l^{\widetilde m}(\eta ,\varphi) =
\sinh ^{l}\eta
e^{-\frac{\alpha_l\eta}{2} }
{\mathcal P}_n^{ \gamma_l , \delta_l }(\coth \eta )e^{i{\widetilde m}\varphi},
\nonumber\\
\gamma_l=\frac{b}{l+\frac{1}{2}} -
\left(l+\frac{1}{2}\right),
 \qquad  \delta_l=-
\frac{b}{
l +\frac{1}{2}
} -\left(
l+\frac{1}{2}\right),\qquad l={\widetilde m}+n.
\label{Jacobi}
\end{gather}
Comparison of (\ref{Jacobi}) to (\ref{decomposition}) allows to conclude on the
existence of f\/inite nonlinear decompositions of the Jacobi polynomials into
associated Legendre functions $(P_l^m)$ according to
\begin{gather*}
\sinh ^{l}\eta
{\mathcal P}_n^{ \gamma_l , \delta_l }(\coth \eta )=
\sum_{m={\widetilde m}}^{m=l}a_{({\widetilde m}+1)(m+1)}^l
P_l^m(\cosh \eta).
%\label{fnt_deco}
\end{gather*}
The latter equation can equally well be used to pin down the
expansion coef\/f\/icients upon using the orthogonality properties of the
associated Legendre functions.

The $({\mathcal C}+2b\coth\eta )$ eigenvalue problem is closely related to
the rigid rotator problem on S$^2$ perturbed by a cotangent
interaction, $({\mathbf L}^2-2b\cot\theta )$.
The two problems have several features in common.
Also the $\cot\theta$ interaction preserves the $(2l+1)$-fold degeneracy
patterns characterizing the spectrum of the free geodesic motion,
despite its non-commutativity with ${\mathbf L}^2$.
The similarities between these two cases are
due to their interrelation by the following complexif\/ications
 \begin{gather}
{\mathcal C}+2b\coth \eta
\stackrel {\eta \to i\theta, b\to -ib}{\longrightarrow}
{\mathbf L}^2 -2b\cot \theta .
\label{cmplxf}
\end{gather}
It is straightforward to prove that in ef\/fect of the complexif\/ications in~(\ref{cmplxf}), the equation~(\ref{bibop1}) is transformed to
\begin{gather}
\left[{\mathbf L}^2-2b\cot\theta \right]
{\mathcal X}_l^{\widetilde{m}}(\theta ,\varphi) =
\left[ e^{-\frac{\alpha_l\eta }{2}}\left( {\mathbf L}^2-
\frac{\alpha_{l}^2}{4}\right)e^{\frac{\alpha_{l}\eta }{2}}\right]
{\mathcal X}_l^{\widetilde{m}}(\theta ,\varphi)\nonumber\\
 \hphantom{\left[{\mathbf L}^2-2b\cot\theta \right]{\mathcal X}_l^{\widetilde{m}}(\theta ,\varphi) = }{}
 =
  \left( \widetilde{\mathbf L}^2 -
\frac{\alpha_l^2}{4}\right){\mathcal X}_l^{\widetilde{m}}(\theta ,\varphi)=
\left( l(l+1) -\frac{\alpha _{(l)}^2}{4}\right)
{\mathcal X}_l^{\widetilde{m}}(\theta, \varphi),
\label{illustr_22}
\end{gather}
 with ${\mathcal X}_l^{\widetilde{m}}(\theta, \varphi)$ being given as
\begin{gather}
{\mathcal X}_l^{\widetilde m}(\theta, \varphi) =
\left[ \sum_{m={\widetilde m}}^{m=l}c_la_{({\widetilde m}+1)(m+1)}^l e^{-im\varphi}
{\widetilde {Y}}_l^{ m}(\theta, \varphi)\right]
e^{i{\widetilde m}\varphi },\nonumber\\
{\widetilde { Y}}_l^m(\theta ,\varphi)  =
e^{-\frac{\alpha_l\theta }{2}} Y_l^m(\theta ,\varphi),
\label{decompositions_1}
\end{gather}
where  $Y_l^m(\theta, \varphi)$ are the standard spherical harmonics, and
the constants $|c_l|=1$ account for possible sign changes in
$a_{({\widetilde m}+1)(m+1)}^l$ in depending on the power of
$(ib)$ contained there. In conclusion, also the cotangent perturbed
rigid rotator can be cast into the form of a Casimir invariant of the
so(3) algebra though in a representation unitarily nonequivalent to the
rotational.

Also for the latter case the exact solutions of equation~(\ref{illustr_22}) are
known~\cite{MolPhys}, and expressed in terms of real Romanovski polynomials~\cite{raposo} as
\begin{gather}
{\mathcal X}_l^{\widetilde m}(\theta, \varphi) =
e^{-\alpha\theta /2} \sin^l\theta R_n^{\frac{2b}{l+\frac{1}{2}}, -\left(l+\frac{1}{2}\right)}(\cot \theta)
e^{i{\widetilde m}\varphi},
\label{par1}
\end{gather}
and also here in comparing (\ref{decompositions_1}) to (\ref{par1})
one f\/inds f\/inite nonlinear decompositions of Romanovski polynomials
into spherical harmonics according to
\begin{gather*}
\sin^l\theta
R_n^{\frac{2b}{l+\frac{1}{2}}, -\left(l+\frac{1}{2}\right)}(\cot \theta)=
\sum_{m={\widetilde m}}^{m=l}c_la_{({\widetilde m}+1)(m+1)}^l
P_l^{ m}(\theta).
%\label{par2}
\end{gather*}

The Romanovski polynomials
satisfy the following dif\/ferential hyper-geometric equation
\begin{gather*}
\big(1+x^2\big)\frac{{\mathrm d}^2R_n^{\alpha, \beta}}{{\mathrm d} x^2}
+2\left(\frac{\alpha }{2} +\beta x
\right)\frac{{\mathrm d}R_n^{\alpha, \beta}}{{\mathrm d}x}
-n(2\beta +n-1)R_n^{\alpha, \beta}=0.
\end{gather*}
They are obtained from the following weight function
\begin{gather*}
\omega ^{\alpha, \beta}(x)=\big(1+x^2\big)^{\beta -1}\exp\big({-}\alpha \cot^{-1}x\big),
\end{gather*}
by means of the Rodrigues formula
\begin{gather*}
R^{\alpha, \beta}_n(x)=\frac{1}{\omega^{\alpha, \beta }(x)}
\frac{{\mathrm d}^n}{{\mathrm d}x^n}
\left[ \big(1+x^2\big)^n\omega^{\alpha, \beta}(x)\right].
\end{gather*}
Although they are related to the Jacobi polynomials as
\begin{gather}
R_n^{\alpha, \beta}(x)=
i^n{\mathcal P}_n^{1-\beta-i\frac{\alpha}{2}, 1-\beta +i\frac{\alpha}{2}}(ix),
\label{Rom_Jac}
\end{gather}
within Bochner's classif\/ication scheme they appear as one of the f\/ive
independent polynomial solutions of the hyper-geometric dif\/ferential equation.
Within this context, equation~(\ref{Rom_Jac})
does not rule out the necessity of considering the Romanovski polynomials,
rather, it presents itself as one out of many possible
interrelationships among polynomials, valid only under certain
restrictions of the parameters of at least one of the involved polynomials.

Another known example of such an interrelationship is
provided by the possibility of
establishing the following link between
(unnormalized) associated Legendre functions $P_l^m$
and Jacobi polynomials
\begin{gather*}
P_l^m(\cos \theta)=\sin ^m\theta {\mathcal P}^{m,m}_{l-m}(\cos \theta).
%\label{Leg_Jac}
\end{gather*}
Obviously, the latter equation does not rule out
the associated Legendre functions as a mathematical entity on its own
in favor of Jacobi polynomials with parameters restricted in this very
particular way.

Back to the $({\mathbf L}^2-2b\cot\theta )$ eigenvalue problem, it
is of interest in the spectroscopy of diatomic molecules and
has been investigated in the work \cite{MolPhys} prior to this,
where one can f\/ind explicit expressions of the expansion
coef\/f\/icients for some of the lowest
$l$ values.
However, there the study has been carried out from a predominantly
spectroscopic, and signif\/icantly subordinate algebraic perspective.
Compared to~\cite{MolPhys}, the present work is entirely focused on
the algebraic aspect of the particle motion on the curved surface
of interest, which parallels the formation mechanism
of the non-unitary similarity transformation connecting
${\mathcal C}$, and $({\mathcal C}+2b\coth \eta)$.
In addition, we wish to emphasize, that casting
the cotangent perturbed motion on S$^2$ in (\ref{illustr_22})
as a Casimir invariant of an intact geometric so(3) algebra,
provides a simple alternative to the deformed
dynamic so(3) Higgs algebra~\cite{Higgs}, which
approaches same problem from the perspective of a~rotational generator on the
plane tangential to the North pole of the sphere, complemented by
the two components of a properly designed
Runge--Lenz vector there.

\subsection{Matrix forms of the wave equations}

Equations~(\ref{bong1}), (\ref{bong}) can be generalized to arbitrary $l$
according to
\begin{gather}
{\mathbf X}_l(\eta, \varphi) =
\begin{pmatrix}
{\mathcal X}_l^0(\eta, \varphi)\\
{\mathcal X}_l^1(\eta ,\varphi)\\
\dots \\
{\mathcal X}_l^l(\eta ,\varphi)
\end{pmatrix}
=
e^{-\frac{\alpha_l \eta }{2}} \begin{pmatrix}
F_l^0(\eta)e^{i0\varphi}\\
F_l^1(\eta)e^{i\varphi }\\
\cdots \\
F_l^l(\eta)e^{il\varphi}
\end{pmatrix} =
A_l(\varphi)
e^{-\frac{\alpha_l \eta }{2}}
\begin{pmatrix}
{\mathcal Y}_l^0(\eta, \varphi )\\
{\mathcal Y}_l^1( \eta, \varphi )\\
\dots \\
{\mathcal Y}_l^l( \eta ,\varphi )
\end{pmatrix}, \nonumber\\
A_l(\varphi) = \begin{pmatrix}
a_{11}^l&a_{12}^le^{-i\varphi} & \dots &a_{1(l+1)}^le^{-il\varphi}\\
0 &a_{22}^l&...&a_{2(l+1)}^le^{-i(l-1)\varphi }\\
\dots &\dots &\dots &\dots \\
0 &0 & 0&a_{(l+1)(l+1)}^l
\end{pmatrix}.
\label{expansions}
\end{gather}
Then the matrix form of equation~(\ref{bibop1}), after accounting for
(\ref{Gl3_Anh}), extends correspondingly to
\begin{gather}
\left[ e^{-\frac{\alpha_l \eta }{2}}
A_l(\varphi){\mathcal C} e^{\frac{\alpha_l \eta }{2}}
A^{-1}_l(\varphi)+\frac{\alpha_l^2}{4}\right]{\mathbf X}_l(\eta,\varphi ),
 = -\epsilon_l {\mathbf X}_l(\eta, \varphi),\nonumber\\
\epsilon_l=-l(l+1)-\frac{\alpha_l^2}{4} =
-l(l+1)-\frac{b^2}{\left(l+\frac{1}{2}\right)^2}.
\label{master}
\end{gather}
In this way,
the eigenvalue problem in equation~(\ref{Gl3}) takes its f\/inal form
\begin{gather}
({\mathcal C} +2b\coth \eta) {\mathbf X}_l(\eta ,\varphi) =
\left[ {\widetilde {\mathcal C}} +\frac{\alpha^2_l}{4}\right]
{\mathbf X}_l(\eta ,\varphi),\nonumber\\
{\widetilde {\mathcal C}} =
e^{-\frac{\alpha_t\eta }{2}}A_l (\varphi) {\mathcal C}
e^{\frac{\alpha_t\eta }{2}}A_l^{-1} (\varphi),
\label{bing_bong}
\end{gather}
with ${\mathcal C}$ being extended to a matrix encoding the dimensionality
of the so(2,1) representation space under consideration.
The explicit matrices $A_l(\varphi)$ for $l=1$ and $l=2$ were given
in equa\-tions~(\ref{bong1}),~(\ref{bong}).
As a reminder, equation (\ref{bing_bong}) can be obtained from
equation~(\ref{insight}) in combination with the recurrence relations in
equations~(\ref{lit})--(\ref{rcr2}).
Equation (\ref{bing_bong}) def\/ines a particular representation of the so(2,1)
algebra
whose eigenvalue problem is related to the standard pseudo-rotational one
by a dilation transformation.
The ${\mathcal {\widetilde C}}$ operator realization is
unitarily nonequivalent to~${\mathcal C}$
and the motion of a particle on~{\bf H}$_+^2$, perturbed by
the~$\coth\eta $ potential,
breaks the pseudo-rotational invariance
at the level of the representation functions.
However, the degeneracy is def\/ined by the eigenvalues of the
Casimir invariant of the algebra and does not depend
on the particular realization of the algebra, a reason for which the
perturbed spectrum in~(\ref{master}) carries same
degeneracy patterns as the unperturbed one, corresponding to~$b=0$.

Finally, obtaining
the counterpart to equation~(\ref{bing_bong})
for the case of the cotangent perturbed rigid rotator on S$^2$
along the line of equations~(\ref{cmplxf})--(\ref{decompositions_1})
is too simple an exercise as to be explicitly worked out here,
and we omit it in favor of a keeping the presentation more concise.

\section{Summary and conclusions}\label{section3}

In the present study, attention was drawn to relevance for physics problems
of non-unitary similarity transformations of some isometry algebras
of curved surfaces. Specif\/ically, the so(1,2) isometry algebra of
the two-dimensional two-sheeted hyperboloid was considered in detail and a~non-unitary similarity
transformation was found that connected
the eigenvalue problems of the Casimir operator of the free geodesic motion
and the one perturbed by a hyperbolic cotangent interaction.

The similarity transformation was concluded from
transparent f\/inite decompositions of the exact
solutions of the Eckart potential
in the bases of exponentially scaled pseudo-spherical harmonics
presented in the above equations~(\ref{bong}), and~(\ref{expansions}).
It furthermore was motivated by the relevance of specif\/ic
recurrence relations (\ref{rcr1})--(\ref{rcr2})
among associated Legendre functions.

A merit of representing the exact solutions of the Eckart potential
as superpositions of damped pseudo-spherical harmonics
is that these expansions would provide a practical tool
in the description of quantum mechanical systems on {\bf H}$_+^2$
whose interactions are only
approximately described by the Eckart potential, in which case
the elements
of the matrices $A_l(\varphi)$ could be considered as parameters to be
adjusted to data.

The perturbed geodesic motion on {\bf H}$_+^2$
in terms of the Eckart potential as considered here,
had the peculiarity that the perturbance respected the degeneracy
of the unperturbed free geodesic motion though it broke the
isometry group symmetry at the level of
the representation functions.
The case under consideration easily extended to the cotangent perturbed
rigid rotator on S$^2$ and verif\/ied similar observations earlier reported in~\cite{MolPhys}.
In this fashion, two examples
of a symmetry breaking at the level of the
representation functions have been constructed, breakdowns that appeared
camouf\/laged by the conservation of the degeneracy patterns in the spectra.

This subtle type of symmetry breaking was visualized in
Figs.~\ref{tld2} and~\ref{baseball} through the
deformation of the metric of the respective {\bf H}$_+^2$ hyperboloid,
and the S$^2$ sphere, by the exponential scalings
$e^{-\alpha\eta/2}|{\mathcal Y}_0^0(\eta,\varphi)$,
and $e^{-\alpha\theta/2}|{\mathcal Y}_0^0(\theta,\varphi)$.

\subsection*{Acknowledgments}
We thank Jose Limon Castillo for constant assistance
in managing computer matters.
Work partly supported by CONACyT-M\'{e}xico under grant number
CB-2006-01/61286.

\pdfbookmark[1]{References}{ref}
\LastPageEnding

\end{document}